# Impact of Newly Measured Nuclear Reaction Rates on $^{26}$Al Ejected Yields from Massive Stars

Umberto Battino [1,*,†], Lorenzo Roberti [2,3,4,†], Thomas V. Lawson [1,5,†], Alison M. Laird [6,†] and Lewis Todd [6]

1. Edward Arthur Milne Centre for Astrophysics, University of Hull, Cottingham Road, Kingston upon Hull HU6 7RX, UK
2. Konkoly Observatory, Research Centre for Astronomy and Earth Sciences, HUN-REN, Konkoly Thege Miklós út 15-17, H-1121 Budapest, Hungary
3. CSFK, MTA Centre of Excellence, Konkoly Thege Miklós út 15-17, H-1121 Budapest, Hungary
4. INAF-Osservatorio Astronomico di Roma Via Frascati 33, I-00040 Monteporzio Catone, Italy
5. Data Science, AI & Modelling Centre (DAIM), University of Hull, Cottingham Road, Kingston upon Hull HU6 7RX, UK
6. School of Physics, Engineering and Technology, University of York, Heslington YO10 5DD, UK
* Correspondence: u.battino@hull.ac.uk
† The NuGrid Collaboration, http://nugrid.github.io.

**Abstract:** Over the last three years, the rates of all the main nuclear reactions involving the destruction and production of $^{26}$Al in stars ($^{26}$Al($n$, $p$)$^{26}$Mg, $^{26}$Al($n$, $\alpha$)$^{23}$Na, $^{26}$Al($p$, $\gamma$)$^{27}$Si and $^{25}$Mg($p$, $\gamma$)$^{26}$Al) have been re-evaluated thanks to new high-precision experimental measurements of their cross-sections at energies of astrophysical interest, considerably reducing the uncertainties in the nuclear physics affecting their nucleosynthesis. We computed the nucleosynthetic yields ejected by the explosion of a high-mass star (20 M$_\odot$, Z = 0.0134) using the FRANEC stellar code, considering two explosion energies, $1.2 \times 10^{51}$ erg and $3 \times 10^{51}$ erg. We quantify the change in the ejected amount of $^{26}$Al and other key species that is predicted when the new rate selection is adopted instead of the reaction rates from the STARLIB nuclear library. Additionally, the ratio of our ejected yields of $^{26}$Al to those of 14 other short-lived radionuclides ($^{36}$Cl, $^{41}$Ca, $^{53}$Mn, $^{60}$Fe, $^{92}$Nb, $^{97}$Tc, $^{98}$Tc, $^{107}$Pd, $^{126}$Sn, $^{129}$I, $^{36}$Cs, $^{146}$Sm, $^{182}$Hf, $^{205}$Pb) are compared to early solar system isotopic ratios, inferred from meteorite measurements. The total ejected $^{26}$Al yields vary by a factor of $\sim$3 when adopting the new rates or the STARLIB rates. Additionally, the new nuclear reaction rates also impact the predicted abundances of short-lived radionuclides in the early solar system relative to $^{26}$Al. However, it is not possible to reproduce all the short-lived radionuclide isotopic ratios with our massive star model alone, unless a second stellar source could be invoked, which must have been active in polluting the pristine solar nebula at a similar time of a core-collapse supernova.

## 1. Introduction

The synthesis occurring in stars (and related explosions) of the radioactive nuclide $^{26}$Al, characterized by a half-life of 0.72 Myr, plays a pivotal role in enhancing our comprehension of the genesis of our solar system and the evolution of stars and galaxies, and is a subject of interest in both $\gamma$-ray astrophysics and cosmochemistry [1,2]. An excess of $^{26}$Mg, the daughter isotope of $^{26}$Al, is found in meteoritic calcium-aluminium-rich inclusions (CAIs), the first solids to form in the protosolar nebula. This observation provides evidence for the injection of live $^{26}$Al during the early stages of the Solar System [3,4]. Consequently, unraveling the origin of $^{26}$Al is imperative for a comprehensive understanding of the birth of the Sun and our planetary system. Additionally, the distinctive $^{26}$Al emission of the diffuse 1809 keV line within our Galaxy, as observed by $\gamma$-ray telescopes [5], serves as a direct tracer of ongoing nucleosynthesis processes enriching the interstellar medium, with its spatial distribution indicating that outflows from massive Wolf-Rayet stars (with masses exceeding 25 $M_\odot$ [6,7]) and core-collapse supernovae are the principal loci for $^{26}$Al production [8], contributing to approximately 70% of the observed live $^{26}$Al in the Milky Way [9,10].

$^{26}$Al is generated during three distinct phases in the life cycle of massive stars: (i) H core burning in Wolf-Rayet stars, characterized by strong mass loss leading to the expulsion of highly $^{26}$Al-enriched layers within the H convective core; (ii) explosive C/Ne burning; and (iii) C/Ne convective shell burning during pre-supernova stages. In the latter phase, the fraction of $^{26}$Al that survives the subsequent explosion is ejected [11,12].

Other contributors to the presence of $^{26}$Al in the Galactic environment include nova explosions [13]. Novae play a role in supplying up to 30% of the live $^{26}$Al observed in the Milky Way [14,15]. Further minor sources are found in Asymptotic Giant Branch (AGB) stars, representing the concluding phase in the life cycle of low- and intermediate-mass stars [16], and Type-I X-ray bursts (assuming enough mass can actually be ejected from the accreting neutron star). In this last case, one of the main channels producing $^{26}$Al is the nuclear reaction chain $^{22}$Mg($\alpha$, $p$)$^{25}$Al($\beta^+$)$^{25}$Mg($p$, $\gamma$)$^{26}$Al, whose reaction rate uncertainty can significantly impact the modeled light curve shape [17]. Across all these sources, $^{26}$Al is generated through the $^{25}$Mg($p$, $\gamma$)$^{26}$Al nuclear reaction, and it is mainly destroyed (apart from its $\beta+$ decay into $^{26}$Mg) by $^{26}$Al($p$, $\gamma$)$^{27}$Si [11] and, if an efficient neutron source is available, by the neutron-induced nuclear reactions $^{26}$Al($n$, $p$)$^{26}$Mg and $^{26}$Al($n$, $\alpha$)$^{23}$Na [18]. The rates of all these key nuclear reactions have been re-evaluated by several authors over the last two years. Both the $^{25}$Mg($p$, $\gamma$)$^{26}$Al and the $^{26}$Al($p$, $\gamma$)$^{27}$Si nuclear reaction rates have been recently re-computed by the authors of Ref. [2], who presented a re-evaluation based on the most recent nuclear data and included the deep underground direct measurement performed by the LUNA collaboration [19]. Additionally, the $^{25}$Mg($p$, $\gamma$)$^{26}$Al nuclear reaction was also re-studied by the authors of Ref. [20], who derived a new reaction rate based on the results of a complete experimental investigation of the 92, 130, and 189 keV resonances (interms of center of mass energy) with the Jinping Underground Nuclear Astrophysics (JUNA) lab. Finally, the authors of Ref. [21] determined a new rate for both the $^{26}$Al($n$, $p$)$^{26}$Mg and $^{26}$Al($n$, $\alpha$)$^{23}$Na nuclear reactions, of which they presented re-evaluations primarily based on high-precision measurements from the n_TOF-CERN facility ([22–24]), supplemented by theoretical calculations and a previous experiment [25] at higher neutron energies, to cover the full range of relevant stellar temperatures.

In this work, we re-compute the ejected nucleosynthetic yields from a 20 $M_\odot$ massive star at solar metallicity including all the new rates of the $^{25}$Mg($p$, $\gamma$)$^{26}$Al, $^{26}$Al($p$, $\gamma$)$^{27}$Si, $^{26}$Al($n$, $p$)$^{26}$Mg and $^{26}$Al($n$, $\alpha$)$^{23}$Na key nuclear reactions involving $^{26}$Al. The main characteristics of our stellar models are presented in Section 2.2. We apply the new rates in the calculation of full stellar and nucleosynthesis models, quantify the impact of the new nuclear reaction rates and compare the results to key observables in Section 3. Our conclusions are presented in Section 4.

## 2. Computational Methods
### 2.1. Simulation Setup

We used the 1D stellar code FRANEC (Frascati Raphson-Newton Evolutionary Code, presented in detail in [26,27]) to calculate the hydrostatic evolution of the models presented in this work. FRANEC solves the equations describing the physical and chemical evolution of a star in hydrostatic and thermal equilibrium, assuming spherical symmetry and applying a classical Raphson-Newton method. The adopted initial composition for the solar metallicity is the one provided by Asplund, 2009 [28] and corresponds to Z = 0.01345. For the simulations, we used an extended nuclear network, already adopted in Roberti et al. [29], including 525 nuclear species extending up to Bi and more than 3000 reactions, fully coupled to the solution of the equations that describe the evolution of the star. The explosion and the associated explosive nucleosyntheses have been computed by means of the HYPERION Lagrangian hydrodynamic flux limited diffusion radiation 1D code (extensively described in [30]). HYPERION solves the full system of hydrodynamic equations governing the supernova evolution by means of the fully Lagrangian scheme of the piecewise parabolic method (PPM), a higher-order extension of Godunov's method, described by [31]. The nuclear network used for the explosion is the same as the one used for the hydrostatic evolution of the star.

## 2.2. Description of the Stellar Models

To evaluate the impact of the newly measured reaction rates on $^{26}$Al, we calculated the evolution of a 20 $M_\odot$ star at solar metallicity adopting three different reaction rate selections for the $^{25}$Mg($p$, $\gamma$)$^{26}$Al, $^{26}$Al($p$, $\gamma$)$^{27}$Si, $^{26}$Al($n$, $p$)$^{26}$Mg and $^{26}$Al($n$, $\alpha$)$^{23}$Na nuclear reaction, as summarised in Table 1. The nuclear reaction rates adopted in the STANDARD case are the same as in [29]; in particular, the $^{25}$Mg($p$, $\gamma$)$^{26}$Al and $^{26}$Al($p$, $\gamma$)$^{27}$Si reaction rates are both from [32], while the $^{26}$Al($n$, $p$)$^{26}$Mg and $^{26}$Al($n$, $\alpha$)$^{23}$Na reaction rates are from [33,34], respectively. In the LA-BA case abbreviation for "LAIRD-BATTINO", the $^{26}$Al($n$, $p$)$^{26}$Mg and $^{26}$Al($n$, $\alpha$)$^{23}$Na reaction rates are both from [21], while the $^{25}$Mg($p$, $\gamma$)$^{26}$Al and $^{26}$Al($p$, $\gamma$)$^{27}$Si reaction rates are both from [2]. Most importantly, we extended both rates beyond 0.7 GK (the highest temperature provided by [2]) using the RATESMC code [32], including higher energy resonances beyond 484 keV based on the information presented in [32]. Our new high-temperature rates are presented in Tables 2–4. Finally, in the JU-LA-BA case (abbreviation for "JUNA-LAIRD-BATTINO") the reaction rates adopted are the same as in the LA-BA case, with the only exception of the $^{25}$Mg($p$, $\gamma$)$^{26}$Al reaction rate which is taken from [20]. Note that we do not attempt any extrapolation of the reaction rate in all the cases where the temperature grid is not extending up to 10 GK, but instead we adopt the value of the last available experimental point.

**Table 1.** Schematic view of the references adopted for the three reaction rate selections presented in this work.

| Nuclear Reaction | STANDARD | JU-LA-BA | LA-BA |
| --- | --- | --- | --- |
| $^{25}$Mg($p$, $\gamma$)$^{26}$Al | Iliadis et al., 2010 [32] | Zhang et al., 2023 [20] | Laird et al., 2023 [2] |
| $^{26}$Al($p$, $\gamma$)$^{27}$Si | Iliadis et al., 2010 [32] | Laird et al., 2023 [2] | Laird et al., 2023 [2] |
| $^{26}$Al($n$, $p$)$^{26}$Mg | Caughlan & Fowler 1988 [33] | Battino et al., 2023 [21] | Battino et al., 2023 [21] |
| $^{26}$Al($n$, $\alpha$)$^{23}$Na | Angulo et al., 1999 [34] | Battino et al., 2023 [21] | Battino et al., 2023 [21] |

**Table 2.** $^{25}$Mg($p$, $\gamma$)$^{26}$Al reaction rates for temperatures higher than 0.7 GK on the ground state of $^{26}$Al in units of [cm$^3$/mol s].

| T[GK] | Lower Limit | Median Rate | Upper Limit |
| --- | --- | --- | --- |
| 0.700 | $1.027 \times 10^2$ | $1.078 \times 10^2$ | $1.136 \times 10^2$ |
| 0.800 | $1.933 \times 10^2$ | $2.022 \times 10^2$ | $2.121 \times 10^2$ |
| 0.900 | $3.195 \times 10^2$ | $3.333 \times 10^2$ | $3.488 \times 10^2$ |
| 1.000 | $4.810 \times 10^2$ | $5.011 \times 10^2$ | $5.240 \times 10^2$ |
| 1.250 | $1.026 \times 10^3$ | $1.068 \times 10^3$ | $1.118 \times 10^3$ |
| 1.500 | $1.740 \times 10^3$ | $1.811 \times 10^3$ | $1.897 \times 10^3$ |
| 1.750 | $2.586 \times 10^3$ | $2.689 \times 10^3$ | $2.815 \times 10^3$ |
| 2.000 | $3.532 \times 10^3$ | $3.667 \times 10^3$ | $3.831 \times 10^3$ |
| 2.500 | $5.615 \times 10^3$ | $5.809 \times 10^3$ | $6.039 \times 10^3$ |
| 3.000 | $7.804 \times 10^3$ | $8.047 \times 10^3$ | $8.329 \times 10^3$ |
| 3.500 | $9.946 \times 10^3$ | $1.023 \times 10^4$ | $1.055 \times 10^4$ |
| 4.000 | $1.193 \times 10^4$ | $1.225 \times 10^4$ | $1.261 \times 10^4$ |
| 5.000 | $1.520 \times 10^4$ | $1.559 \times 10^4$ | $1.600 \times 10^4$ |
| 6.000 | $1.748 \times 10^4$ | $1.792 \times 10^4$ | $1.839 \times 10^4$ |
| 7.000 | $1.890 \times 10^4$ | $1.938 \times 10^4$ | $1.989 \times 10^4$ |
| 8.000 | $1.966 \times 10^4$ | $2.017 \times 10^4$ | $2.071 \times 10^4$ |
| 9.000 | $1.994 \times 10^4$ | $2.047 \times 10^4$ | $2.103 \times 10^4$ |
| 10.000 | $1.989 \times 10^4$ | $2.043 \times 10^4$ | $2.100 \times 10^4$ |

**Table 3.** $^{25}$Mg($p, \gamma$)$^{26}$Al reaction rates for temperatures higher than 0.7 GK on the isomeric state of $^{26}$Al in units of [cm$^3$/mol s].

| T[GK] | Lower Limit | Median Rate | Upper Limit |
|---|---|---|---|
| 0.700 | $2.887 \times 10^1$ | $3.054 \times 10^1$ | $3.237 \times 10^1$ |
| 0.800 | $5.899 \times 10^1$ | $6.232 \times 10^1$ | $6.610 \times 10^1$ |
| 0.900 | $1.043 \times 10^2$ | $1.102 \times 10^2$ | $1.171 \times 10^2$ |
| 1.000 | $1.661 \times 10^2$ | $1.756 \times 10^2$ | $1.870 \times 10^2$ |
| 1.250 | $3.944 \times 10^2$ | $4.169 \times 10^2$ | $4.451 \times 10^2$ |
| 1.500 | $7.198 \times 10^2$ | $7.593 \times 10^2$ | $8.101 \times 10^2$ |
| 1.750 | $1.125 \times 10^3$ | $1.183 \times 10^3$ | $1.259 \times 10^3$ |
| 2.000 | $1.592 \times 10^3$ | $1.668 \times 10^3$ | $1.767 \times 10^3$ |
| 2.500 | $2.631 \times 10^3$ | $2.741 \times 10^3$ | $2.881 \times 10^3$ |
| 3.000 | $3.705 \times 10^3$ | $3.843 \times 10^3$ | $4.016 \times 10^3$ |
| 3.500 | $4.720 \times 10^3$ | $4.883 \times 10^3$ | $5.079 \times 10^3$ |
| 4.000 | $5.626 \times 10^3$ | $5.807 \times 10^3$ | $6.020 \times 10^3$ |
| 5.000 | $7.033 \times 10^3$ | $7.240 \times 10^3$ | $7.474 \times 10^3$ |
| 6.000 | $7.930 \times 10^3$ | $8.154 \times 10^3$ | $8.400 \times 10^3$ |
| 7.000 | $8.428 \times 10^3$ | $8.660 \times 10^3$ | $8.910 \times 10^3$ |
| 8.000 | $8.641 \times 10^3$ | $8.873 \times 10^3$ | $9.124 \times 10^3$ |
| 9.000 | $8.660 \times 10^3$ | $8.890 \times 10^3$ | $9.135 \times 10^3$ |
| 10.000 | $8.550 \times 10^3$ | $8.776 \times 10^3$ | $9.015 \times 10^3$ |

**Table 4.** $^{26}$Al($p, \gamma$)$^{27}$Si reaction rates for temperatures higher than 0.7 GK on the ground state of $^{26}$Al in units of [cm$^3$/mol s].

| T[GK] | Lower Limit | Median Rate | Upper Limit |
|---|---|---|---|
| 0.700 | $4.640 \times 10^1$ | $5.050 \times 10^1$ | $5.500 \times 10^1$ |
| 0.800 | $8.020 \times 10^1$ | $8.730 \times 10^1$ | $9.520 \times 10^1$ |
| 0.900 | $1.220 \times 10^2$ | $1.330 \times 10^2$ | $1.450 \times 10^2$ |
| 1.000 | $1.710 \times 10^2$ | $1.860 \times 10^2$ | $2.020 \times 10^2$ |
| 1.250 | $3.170 \times 10^2$ | $3.420 \times 10^2$ | $3.700 \times 10^2$ |
| 1.500 | $4.880 \times 10^2$ | $5.230 \times 10^2$ | $5.610 \times 10^2$ |
| 1.750 | $6.730 \times 10^2$ | $7.160 \times 10^2$ | $7.640 \times 10^2$ |
| 2.000 | $8.600 \times 10^2$ | $9.120 \times 10^2$ | $9.680 \times 10^2$ |
| 2.500 | $1.210 \times 10^3$ | $1.280 \times 10^3$ | $1.350 \times 10^3$ |
| 3.000 | $1.500 \times 10^3$ | $1.580 \times 10^3$ | $1.670 \times 10^3$ |
| 3.500 | $1.720 \times 10^3$ | $1.820 \times 10^3$ | $1.920 \times 10^3$ |
| 4.000 | $1.880 \times 10^3$ | $1.990 \times 10^3$ | $2.100 \times 10^3$ |
| 5.000 | $2.040 \times 10^3$ | $2.160 \times 10^3$ | $2.300 \times 10^3$ |
| 6.000 | $2.050 \times 10^3$ | $2.170 \times 10^3$ | $2.320 \times 10^3$ |
| 7.000 | $1.960 \times 10^3$ | $2.080 \times 10^3$ | $2.210 \times 10^3$ |
| 8.000 | $1.830 \times 10^3$ | $1.930 \times 10^3$ | $2.050 \times 10^3$ |
| 9.000 | $1.680 \times 10^3$ | $1.780 \times 10^3$ | $1.880 \times 10^3$ |
| 10.000 | $1.550 \times 10^3$ | $1.630 \times 10^3$ | $1.730 \times 10^3$ |

In this work, we focused on non-magnetic and non-rotating massive star models simulations, as we plan to extend our analysis to rotating models in a follow-up study. The physics inputs and assumptions are the same as used in [27,29]; in the following, we briefly recall the main relevant features. The convective zones are treated using the mixing length theory (MLT), and their borders are defined according to the Ledoux criterion in H-burning regions and to the Schwarzschild criterion elsewhere. Semiconvection is taken into account, as discussed in [35]. The MLT and semiconvection parameters are $\alpha_{\text{MLT}}$ = 2.1 and $\alpha_{\text{semi}}$ = 0.02, respectively. The convective boundary mixing (CBM), implemented as 0.2 $H_P$ of overshooting, is considered only at the edge of the H convective core during the main sequence (MS). Mass loss has been included following the prescriptions of [36,37] for the Blue Super Giant (BSG) phase, [38,39] for the Red Super Giant (RSG) phase, and [40] for the Wolf-Rayet phase (WR). We also included the dynamical mass loss caused by the approach of the luminosity of the star to the Eddington limit by removing all the zones where the Eddington factor $\Gamma = L_{\rm rad}/L_{\rm EDD} > 1$. The explosion is triggered through a thermal bomb, i.e., by depositing a certain amount of thermal energy at the edge of the Fe core in the pre-supernova model. The injected energy is such that to achieve a final kinetic energy of the ejecta (measured at infinity) of $1.2 \times 10^{51}$ erg, the typical kinetic energy assumed for SN 1987A and widely used in the literature [41–44] is employed. The final mass-cut (that defines the remnant mass) is determined by the total amount of fallback $\sim 10^7$ s after the core-collapse and corresponds to 2.58 $M_\odot$, in agreement with other 20 $M_\odot$ CCSN models available in literature [12,45,46]. We calculate an additional CCSN model, imposing a final kinetic energy of $3 \times 10^{51}$ erg, to study the effect of a more energetic explosion on the production of $^{26}$Al. In this case, the explosion energy is high enough to completely eject all the layers above the Fe core. The mass-cut is therefore chosen so that the ejecta contain 0.07 $M_\odot$ of $^{56}$Ni, to match the amount of $^{56}$Ni estimated from SN 1987A, and it is 1.86 $M_\odot$. Finally, we did not include any "*mixing and fall back*" prescriptions in any of our CCSN simulations (see, e.g., [47,48]). A summary of the main properties of the reference model is reported in Table 5. The final structure is instead shown in Figure 1. We remind the reader that the explosive nucleosynthesis depends only on the temperature and density in the post-shock zones, and we use the following definition (e.g., [49–51]):

- Complete Si burning (Si$_c$): $T_{\rm shock} > 5$ GK;
- Incomplete Si burning (Si$_i$): 5 GK $> T_{\rm shock} > 4$ GK;
- Explosive O burning (O): 4 GK $> T_{\rm shock} > 3.3$ GK;
- Explosive Ne burning (Ne): 3.3 GK $> T_{\rm shock} > 2.1$ GK;
- Explosive C burning (C): 2.1 GK $> T_{\rm shock} > 1.9$ GK.

**Table 5.** Summary of the key characteristics of the presupernova (PSN) and explosive models presented in this work.

| | |
|---|---|
| Mass-loss scheme | Vink [36,37], de Jaeger [38], Nugis [40], van Loon [39] |
| Convection criteria | Ledoux in H burning zones, Schwarzschild elsewhere |
| $\alpha_{\rm MLT}$ | 2.1 |
| $\alpha_{\rm semi}$ | 0.02 |
| CBM | 0.2 $H_{\rm P}$ of overshooting (H burning) |
| Initial mass | 20 $M_\odot$ |
| Initial metallicity | $1.345 \times 10^{-2}$ (Asplund 2009 [28]) |
| Initial H mass fraction | 0.721 |
| Initial $^4$He mass fraction | 0.265 |
| $^{12}$C mass fraction (He exhaustion) | 0.286 |
| CO core (He exhaustion) | 4.91 $M_\odot$ |
| Lifetime (PSN) | 9.87 Myr |
| Total Mass (PSN) | 7.46 $M_\odot$ |
| log $T_{\rm eff}$ (PSN) | 4.30 |
| log $L/L_\odot$ (PSN) | 5.31 |
| Explosion scheme | Thermal bomb [30] |
| Thermal energy injected | $2.61 \times 10^{51}$ erg, $4.41 \times 10^{51}$ erg |
| Explosion energy (at infinity) | $1.20 \times 10^{51}$ erg, $3.00 \times 10^{51}$ erg |
| Remnant mass | 2.58 $M_\odot$, 1.86 $M_\odot$ |

$^{26}$Al in CCSNe is usually produced between explosive Ne and C zones. In our models, $^{26}$Al is synthesized at temperatures between 2.60 and 1.74 GK. It is clearly visible in Figure 2, which shows the $^{25}$Mg(p, γ)$^{26}$Al nuclear reaction rate ratio between [2,20], how, in this temperature range, the rate from [2] is very close to the rate from [20], while it is between two and three factors lower at the lowest temperatures. This is mainly due to the shifted resonance energy computed when taking the atomic binding into account, as described in [2], while this effect was not included by [20].

Below $T_{shock}$ = 1.9 GK, the temperature is too low on the typical explosion timescale (∼1 s) to activate other major explosive burning stages, and therefore, the matter is ejected without any further reprocessing. However, some secondary processes may still occur, such as the neutron burst in the He shell, required to explain Mo and Zr anomalies in pre-solar silicon carbide grains [52].

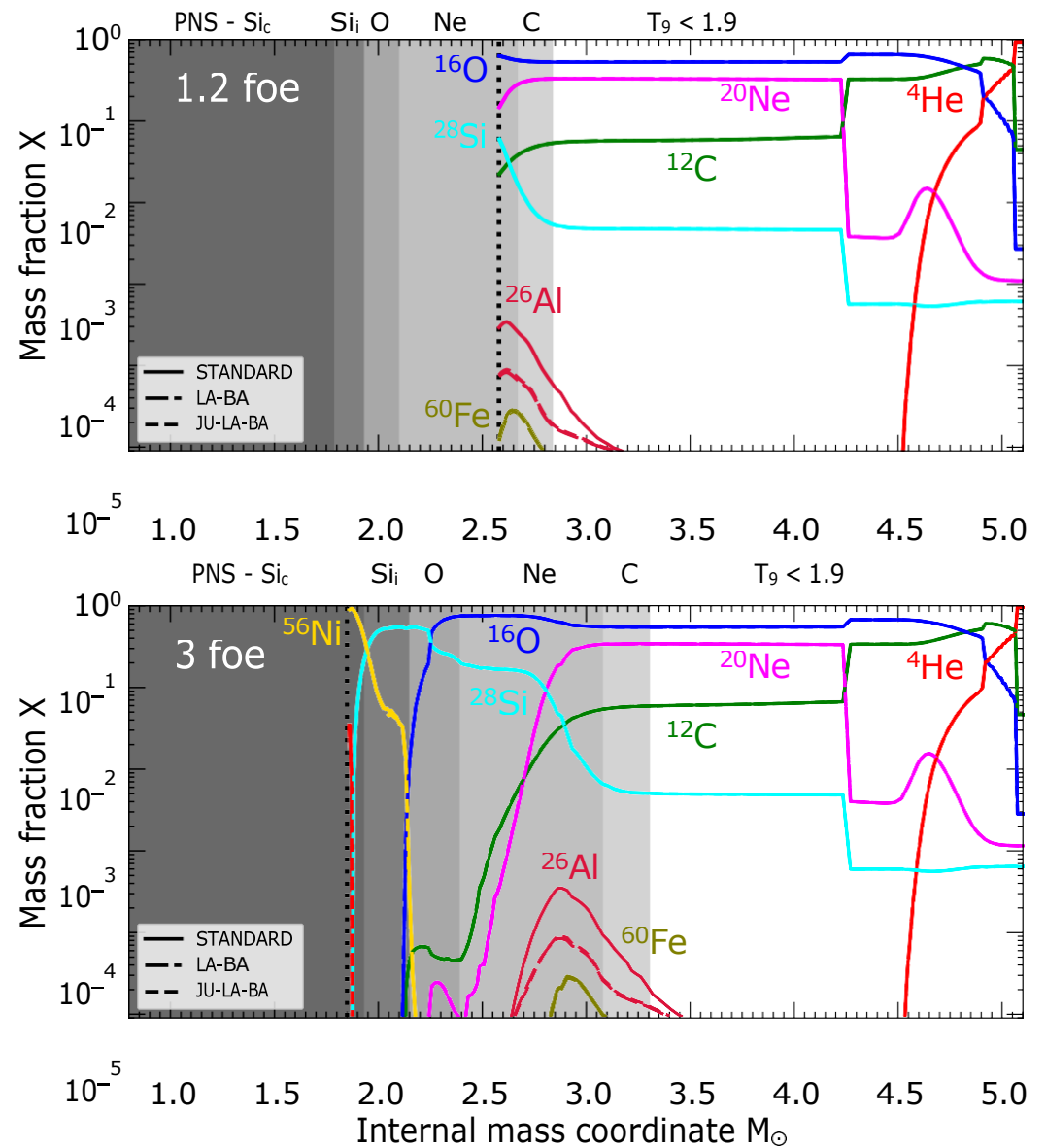

**Figure 1.** Abundances in mass fraction of key nuclear species as a function of the internal mass coordinate in the CCSN models exploding with $1.2 \times 10^{51}$ erg (**upper** panel) and $3 \times 10^{51}$ erg (**lower** panel). The gray shaded areas represent each explosive burning stage (see text); the vertical dotted line identifies the location of the mass-cut. The zone below the mass-cut contains the proto-neutron star (PNS), plus the fallback of the zone exposed to the explosive nucleosynthesis, depending on the explosion energy.

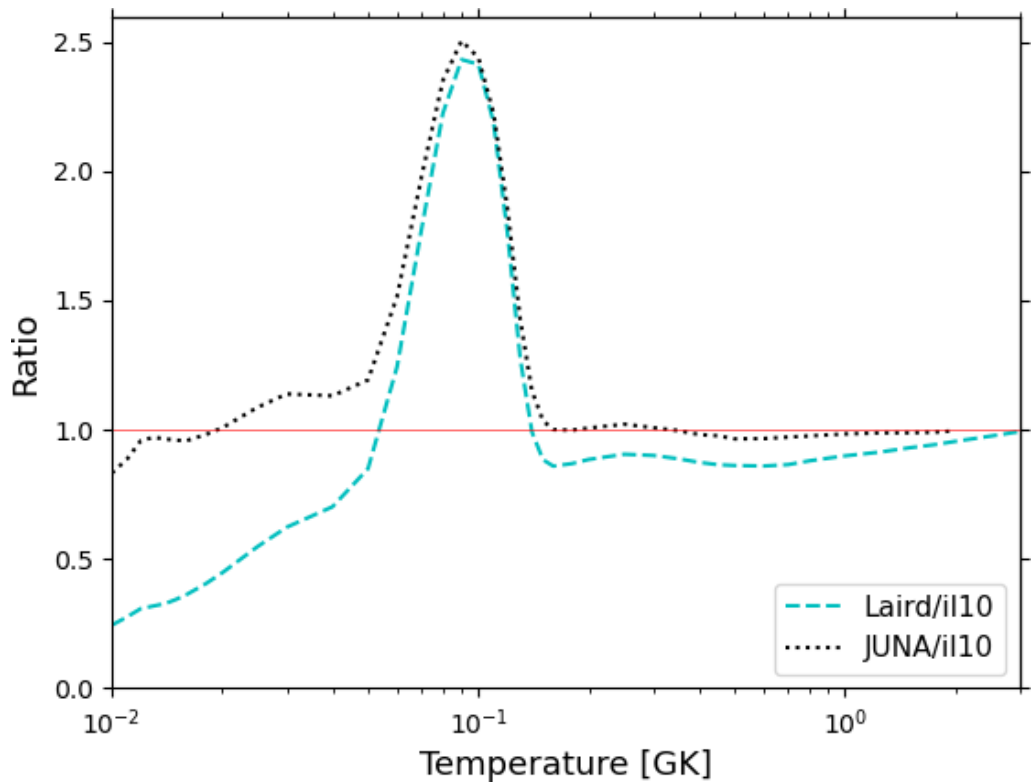

**Figure 2.** Comparison of [2,20] vs. [32] $^{25}$Mg(p, γ)$^{26}$Al nuclear reaction rate ratio as a function of temperature.

## 3. Results

### 3.1. Impact of New Nuclear Reaction Rates on the Ejected Yields

We re-computed the complete nucleosynthesis fully coupled to the structural evolution of our massive star model, using the three network settings described in Section 2.2. The resulting ejected yields for Ne, Na, Mg, Al and Si stable isotopes and $^{26}$Al and $^{60}$Fe (i.e., the species most affected by the newly measured reaction rates we test in this work) are presented in Table 6. Only minor differences are present between the LA-BA and JU-LA-BA cases. Indeed, the total rate of the $^{25}$Mg($p$, γ)$^{26}$Al between 2.60 and 1.74 GK from Laird et al., 2023 [2] is very close to the rate from JUNA [20], differing on average by about 4%. Since the only difference between the JU-LA-BA and LA-BA cases is the different $^{25}$Mg($p$, γ)$^{26}$Al reaction rate, this causes the small difference of the same order between the two cases in the final $^{26}$Al abundance, being 4% lower in LA-BA compared to JU-LA-BA. Larger differences are instead present between the STANDARD case and both the JU-LA-BA and LA-BA cases, due to the different $^{26}$Al($n$, α)$^{23}$Na and $^{26}$Al($n$, $p$)$^{26}$Mg reaction rates adopted. In particular, the $^{26}$Al ejected yields change by a factor between 2 and 3 depending on the explosion energy. This is consistent with the results of [21], who noticed how the final abundance of $^{26}$Al produced in their supernova simulations was about a factor of 2 lower when their $^{26}$Al($n$, α)$^{23}$Na and $^{26}$Al($n$, $p$)$^{26}$Mg reaction rates were adopted (as in JU-LA-BA and LA-BA) instead of those from [33,34] (as in STANDARD).

**Table 6.** Total explosive ejected yields (in $M_\odot$) of $^{26}$Al and other key species of our models for different selections of nuclear reaction rates (see main text for more details).

| Species | STANDARD | JU-LA-BA | LA-BA |
|---|---|---|---|
| $1.2 \times 10^{51}$ erg | | | |
| $^{20}$Ne | $5.60 \times 10^{-1}$ | $5.60 \times 10^{-1}$ | $5.60 \times 10^{-1}$ |
| $^{23}$Na | $1.07 \times 10^{-2}$ | $1.07 \times 10^{-2}$ | $1.07 \times 10^{-2}$ |
| $^{24}$Mg | $9.32 \times 10^{-2}$ | $9.32 \times 10^{-2}$ | $9.33 \times 10^{-2}$ |
| $^{25}$Mg | $1.66 \times 10^{-2}$ | $1.66 \times 10^{-2}$ | $1.66 \times 10^{-2}$ |
| $^{26}$Mg | $1.65 \times 10^{-2}$ | $1.65 \times 10^{-2}$ | $1.65 \times 10^{-2}$ |
| $^{26}$Al | $7.02 \times 10^{-5}$ | $2.75 \times 10^{-5}$ | $2.65 \times 10^{-5}$ |
| $^{27}$Al | $1.14 \times 10^{-2}$ | $1.13 \times 10^{-2}$ | $1.14 \times 10^{-2}$ |
| $^{28}$Si | $2.13 \times 10^{-2}$ | $2.14 \times 10^{-2}$ | $2.13 \times 10^{-2}$ |
| $^{29}$Si | $3.17 \times 10^{-3}$ | $3.17 \times 10^{-3}$ | $3.18 \times 10^{-3}$ |
| $^{30}$Si | $1.96 \times 10^{-3}$ | $1.94 \times 10^{-3}$ | $1.95 \times 10^{-3}$ |
| $^{60}$Fe | $1.19 \times 10^{-5}$ | $1.19 \times 10^{-5}$ | $1.17 \times 10^{-5}$ |
| $3 \times 10^{51}$ erg | | | |
| $^{20}$Ne | $4.79 \times 10^{-1}$ | $4.79 \times 10^{-1}$ | $4.79 \times 10^{-1}$ |
| $^{23}$Na | $8.80 \times 10^{-3}$ | $8.79 \times 10^{-3}$ | $8.79 \times 10^{-3}$ |
| $^{24}$Mg | $9.81 \times 10^{-2}$ | $9.78 \times 10^{-2}$ | $9.82 \times 10^{-2}$ |
| $^{25}$Mg | $1.44 \times 10^{-2}$ | $1.44 \times 10^{-2}$ | $1.44 \times 10^{-2}$ |
| $^{26}$Mg | $1.44 \times 10^{-2}$ | $1.44 \times 10^{-2}$ | $1.44 \times 10^{-2}$ |
| $^{26}$Al | $9.68 \times 10^{-5}$ | $3.36 \times 10^{-5}$ | $3.26 \times 10^{-5}$ |
| $^{27}$Al | $1.20 \times 10^{-2}$ | $1.19 \times 10^{-2}$ | $1.20 \times 10^{-2}$ |
| $^{28}$Si | $2.82 \times 10^{-1}$ | $2.82 \times 10^{-1}$ | $2.82 \times 10^{-1}$ |
| $^{29}$Si | $4.99 \times 10^{-3}$ | $5.03 \times 10^{-3}$ | $5.03 \times 10^{-3}$ |
| $^{30}$Si | $6.94 \times 10^{-3}$ | $6.88 \times 10^{-3}$ | $6.89 \times 10^{-3}$ |
| $^{60}$Fe | $1.14 \times 10^{-5}$ | $1.14 \times 10^{-5}$ | $1.12 \times 10^{-5}$ |

*3.2. Comparison to Ess Ratios*

The early Solar System (ESS) was enriched in short-lived radionuclides (SLRs, with half-lives ranging between 0.1 and 100 Myr) at the time of its formation 4.6 Gyrs ago [4]. This evidence is gleaned from the examination of meteorite compositions, wherein excesses of daughter isotopes originating from radioactive nuclei are detected. While the majority of SLRs possess half-lives long enough (exceeding approximately 2 Myr) to be adequately explained by their production in the Galaxy [53], isotopes with shorter half-lives (such as $^{26}$Al, $^{36}$Cl and $^{41}$Ca) defy explanation through galactic chemical evolution (GCE) processes alone. Consequently, the presence and abundance of these isotopes necessitate one or more additional local stellar sources, potentially originating from the same molecular cloud as the Sun. These local sources would have directly enriched the protosolar nebula from which the Sun formed or the pre-existing proto-planetary disk (for comprehensive reviews of various scenarios, see, e.g., [4]).

We compared the yields of each of our stellar model to ESS abundances, as outlined in [4] and Lawson et al. (in prep). In order to probe the formation of the Solar System, we consider isotopes that have half-lives comparable to the time frame of the formation of the Solar System from its molecular cloud (approx. 15 Myrs, [54]), and compare their abundances with the ejected yields from our models. The aim is to test the hypothesis that a core-collapse supernova event could have produced all the SLRs in the proportion needed to reproduce their ESS abundances measured in calcium-aluminium-rich inclusions (CAIs, the first solids to have formed in the protosolar nebula), hence shaping the environment for the birth of the Sun. SLR daughter and reference signatures can be detected in meteoritic samples and are further used to estimate the abundances of SLRs in the ESS [4,55]. A list of these SLRs, daughter isotopes, reference isotopes, half-lives, mean lifetime and ESS abundance ratios can be found in Table 7. The nucleosynthesis of the SLRs in both massive star and CCSN are detailed in [12].

In order to calculate the contribution of ejecta from a model to the ESS, we must take into account the effect of dilution and radioactive decay. We can calculate a dilution factor ($f$) for each model, defined by

$$f = \frac{M_{SLR}^{ESS}}{M_{SLR}^{*}} \tag{1}$$

Here, $M_{SLR}^{ESS}$ is the observed abundance of a reference SLR and $M_{SLR}^{*}$ is the total ejected mass of the SLR from the model.

**Table 7.** The full list of SLRs considered in this work, listed with reference isotopes, $T_{1/2}$, $\tau$, and ESS ratios. Data gathered from [4]. If no error is given, the value is an upper limit.

| SLR | Daughter | Reference | $T_{1/2}$ (yr) | $\tau$ (yr) | ESS Ratio |
|---|---|---|---|---|---|
| $^{26}$Al | $^{26}$Mg | $^{27}$Al | $7.170 \times 10^5$ | $1.034 \times 10^6$ | $(5.23 \pm 0.13) \times 10^{-5}$ |
| $^{36}$Cl | $^{36}$S | $^{35}$Cl | $3.010 \times 10^5$ | $4.343 \times 10^5$ | $(2.44 \pm 0.65) \times 10^{-5}$ |
| $^{41}$Ca | $^{41}$K | $^{40}$Ca | $9.940 \times 10^4$ | $1.434 \times 10^5$ | $(4.6 \pm 1.9) \times 10^{-9}$ |
| $^{53}$Mn | $^{53}$Cr | $^{55}$Mn | $3.740 \times 10^6$ | $5.396 \times 10^6$ | $(7 \pm 1) \times 10^{-6}$ |
| $^{60}$Fe | $^{60}$Ni | $^{56}$Fe | $2.620 \times 10^6$ | $3.780 \times 10^6$ | $(1.01 \pm 0.27) \times 10^{-8}$ |
| $^{92}$Nb | $^{92}$Zr | $^{92}$Mo | $3.470 \times 10^7$ | $5.006 \times 10^7$ | $(3.2 \pm 0.3) \times 10^{-5}$ |
| $^{97}$Tc | $^{97}$Mo | $^{98}$Ru | $4.210 \times 10^6$ | $6.074 \times 10^6$ | $<1.1 \times 10^{-5}$ |
| $^{98}$Tc | $^{98}$Ru | $^{98}$Ru | $4.200 \times 10^6$ | $6.059 \times 10^6$ | $<6 \times 10^{-5}$ |
| $^{107}$Pd | $^{107}$Ag | $^{108}$Pd | $6.500 \times 10^6$ | $9.378 \times 10^6$ | $(6.6 \pm 0.4) \times 10^{-5}$ |
| $^{126}$Sn | $^{126}$Te | $^{124}$Sn | $2.300 \times 10^5$ | $3.318 \times 10^5$ | $<3 \times 10^{-6}$ |
| $^{129}$I | $^{129}$Xe | $^{127}$I | $1.570 \times 10^7$ | $2.265 \times 10^7$ | $(1.28 \pm 0.03) \times 10^{-4}$ |
| $^{135}$Cs | $^{135}$Ba | $^{133}$Cs | $2.300 \times 10^6$ | $3.318 \times 10^6$ | $<2.8 \times 10^{-6}$ |
| $^{146}$Sm | $^{142}$Nd | $^{144}$Sm | $6.800 \times 10^7$ | $9.810 \times 10^7$ | $(8.28 \pm 0.44) \times 10^{-3}$ |
| $^{182}$Hf | $^{182}$W | $^{180}$Hf | $8.900 \times 10^6$ | $1.284 \times 10^7$ | $(1.018 \pm 0.043) \times 10^{-4}$ |
| $^{205}$Pb | $^{205}$Tl | $^{204}$Pb | $1.730 \times 10^7$ | $2.496 \times 10^7$ | $(1.8 \pm 1.2) \times 10^{-3}$ |

We use $^{26}$Al to define the $f$ for each model ejecta, as the $^{26}$Al/$^{27}$Al ratio is very well established [56,57]. A time delay is then applied to the ejecta to account for the time between the ejection of the SLRs by the stellar source and its incorporation into the CAIs. The decrease in the abundance of the radioactive species is then calculated using the radioactive decay equation

$$M_{SLR}(t) = M_{SLR}(t_0) e^{-\frac{ln(2)\Delta t}{T_{1/2}}} \tag{2}$$

In this equation, the initial mass of the isotope ($M_{SLR}(t_0)$) is decayed by the half-life of the SLR ($T_{1/2}$). This provides the final decayed abundance of the ejecta ($M_{SLR}(t)$) after a time delay.

The SLR yields of each model are decayed using a time delay of 2.5 Myr. This is a good representative value [4], determined by examining the decay profiles of each SLR (with a dilution $f$ calculated using $^{26}$Al, Lawson et al. in prep).

In Figure 3, we present the ESS abundance predicted by our $1.2 \times 10^{51}$ erg explosive model for each SLR once decayed using Equation (2) and diluted using $^{26}$Al to calculate $f$. In order to fully replicate the ESS abundances, each isotope should lie on (or potentially below) the dashed threshold line. Any SLR above this threshold are overproduced and require an additional $^{26}$Al source. SLRs below the threshold may be polluted by an additional event or through galactic chemical evolution (GCE) contributions. We separate the top and bottom panels of Figure 3 into isotopes with lower/equal and greater half-lives than $^{60}$Fe, respectively. Isotopes in the top panel have the shortest half-lives, and therefore, their ratio to $^{26}$Al should ideally lie very close to the ESS value, since there would not be enough time for GCE to fill the gap in case of very low values (which can instead play a role for the longer-lived isotopes in the bottom panel). The models presented in this work are plotted as single points (a blue diamond for the standard model, blue hexagon for JU-LA-BA and a red star for LA-BA) and are compared to the 23 models with initial mass 20 $M_\odot$ of [12] (covering an explosion range of 0.53–8.86×$10^{51}$ erg and remnant mass range of 1.74–3.40 $M_\odot$), which are plotted as box-plots and represent the impact of different stellar physics inputs. As the LA-BA and JU-LA-BA models have less $^{26}$Al, they have an increased $M_{model}/M_{ESS}$ across all isotopes.

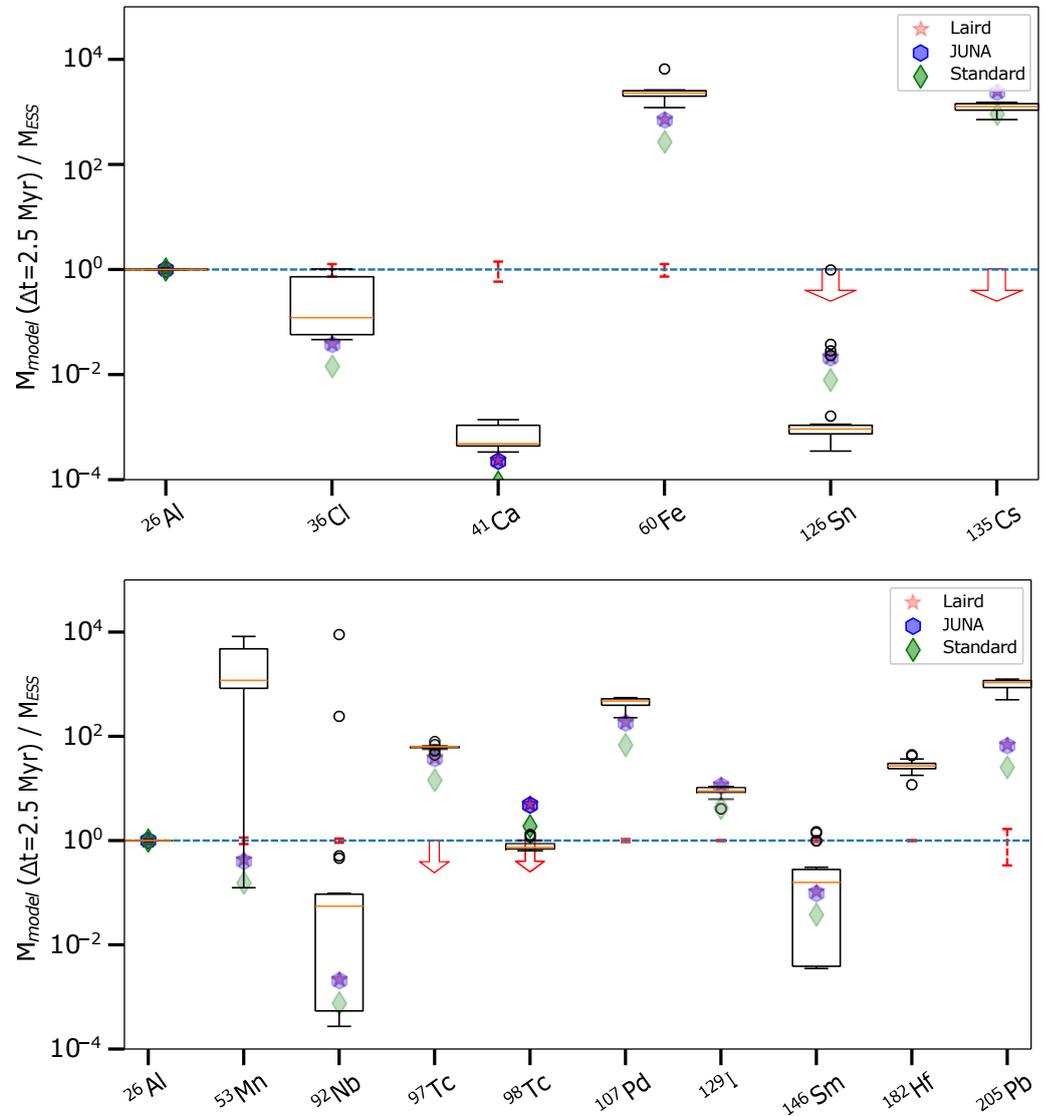

**Figure 3.** Ratio of decayed model ejecta, $M_{model}$, and ESS observation, $M_{ESS}$, for each isotope examined. Within these figures, the horizontal dashed line represents the threshold value of 1. The red error bars represent the error in the ESS ratio of the SLR in question; if the ESS ratio is given as an upper limit, a downward arrow is used. The isolation time for each model is given in the x-axis label as $\Delta t$, and $f$ is calculated using $^{26}$Al. Box-plots in these plots are the 20 $M_\odot$ ejecta of [12]. The upper panel shows isotopes with $T_{1/2} \leq 2.620 \times 10^6$ yr (the $T_{1/2}$ of $^{60}$Fe), and the lower panel shows isotopes with $T_{1/2} \geq 2.620 \times 10^6$ yr. In line with the definition of a *box-plot*, the box size shows the range of values between the 25th and the 75th percentile of all the models in the dataset. The error bars on the box-plot denote the lower (upper) quartiles and are given by subtracting (adding) to the position of the 25th (75th) percentile 1.5 times the interquartile range (i.e., the box size). All values above (below) these thresholds are plotted as outliers. Outlying models outside the outer quartiles are shown as small circles for both decayed and undecayed results, using the same colour scheme as the boxes.

Starting with the isotopes with half-lives less than 2.62 Myr (the $T_{1/2}$ of $^{60}$Fe, top panel of Figure 3), $^{36}$Cl is underproduced in the stellar models presented in this work. However, considering the range of $^{36}$Cl/$^{26}$Al covered by [12] models, it might be in agreement with the observed ESS values if different explosion energies are considered. $^{41}$Ca and $^{126}$Sn are both underproduced, even considering different stellar physics inputs. A similar case is represented by $^{60}$Fe and $^{135}$Cs, which are overproduced in both our models and in those by [12]. Considering the isotopes with half-lives longer than 2.62 Myr (bottom panel of Figure 3), $^{53}$Mn, and $^{98}$Tc are in agreement within nuclear and stellar uncertainties with their ESS values, while $^{92}$Nb and $^{146}$Sm are underproduced, but could be brought into agreement with ESS values by GCE contributions. On the other hand, $^{97}$Tc, $^{107}$Pd, $^{129}$I, $^{182}$Hf and $^{205}$Pb are all overproduced.

Therefore, we notice how, even considering the most recent and up-to-date nuclear physics input, only 5 out of the 14 SLRs considered here are consistent with their observed ESS values. Two potential solutions could be represented by either a different astrophysical scenario that is able to perform better against observations, or an additional pollution event producing more $^{26}$Al and less of the overproduced SLRs (such as $^{60}$Fe) that happened close in time (within about 2.5 Myr) and space to a core-collapse supernova.

## 4. Conclusions

In this work, we computed the evolution of a high-mass star (20 $M_\odot$, Z = 0.01345) and the nucleosynthetic yields ejected by its explosion at two different energies, $1.2 \times 10^{51}$ erg and $3 \times 10^{51}$ erg. We included all the updated rates of the relevant nuclear reactions for $^{26}$Al nucleosynthesis, i.e., $^{26}$Al($n$, $p$)$^{26}$Mg, $^{26}$Al($n$, $\alpha$)$^{23}$Na, $^{26}$Al($p$, $\gamma$)$^{27}$Si and $^{25}$Mg($p$, $\gamma$)$^{26}$Al.

We found a substantial decrease in the ejected amount of $^{26}$Al in the JU-LA-BA and LA-BA cases, between a factor of two and three, compared to the STANDARD case. This is consistent with the impact of the newest $^{26}$Al($n$, $\alpha$)$^{23}$Na and $^{26}$Al($n$, $p$)$^{26}$Mg reaction rates found by [21].

We then compared the ejected yields of our models for 14 SLRs ($^{36}$Cl, $^{41}$Ca, $^{53}$Mn, $^{60}$Fe, $^{92}$Nb, $^{97}$Tc, $^{98}$Tc, $^{107}$Pd, $^{126}$Sn, $^{129}$I, $^{36}$Cs, $^{146}$Sm, $^{182}$Hf, $^{205}$Pb) with their abundances in the ESS, in order to check if a core-collapse supernova could represent a realistic scenario for the circumstances and the environment of the birth of the Sun. Only 5 out of the 14 SLRs are consistent with their observed ESS abundances. At least two potential solutions should be considered: (1) A different astrophysical scenario that is able to perform better against observations. One possibility may be represented by rotating Wolf-Rayet stars, where the rotation-enhanced mass loss may limit the amount of hydrogen to be converted into $^{14}$N during the core H-burning phase and hence the amount of neutrons released by the $^{22}$Ne($\alpha$, $n$)$^{25}$Mg during the He-burning phase (since $^{22}$Ne is produced by a sequence of $\alpha$-captures on $^{14}$N). This may therefore limit the production of n-capture SLRs like $^{60}$Fe and $^{135}$Cs. (2) An additional pollution event producing more $^{26}$Al and less of the overproduced SLRs (such as $^{60}$Fe) that happened close in time (within about 2.5 Myr) and space to a core-collapse supernova.

**Author Contributions:** Conceptualization, U.B.; methodology, U.B.; software, U.B., L.R. and T.V.L.; validation, U.B. and T.V.L.; formal analysis, U.B., L.R. and T.V.L.; investigation, U.B.; resources, U.B.; data curation, U.B., L.R. and T.V.L.; writing—original draft preparation, U.B., L.R. and T.V.L.; writing—review and editing, T.V.L., A.M.L. and L.T.; visualization, U.B., L.R., T.V.L. and L.T.; supervision, A.M.L.; project administration, U.B.; funding acquisition, L.R. All authors have read and agreed to the published version of the manuscript.

**Funding:** This work was supported by the European Union (ChETEC-INFRA, project No. 101008324). L.R. thanks the support from the Lendület Program LP2023-10 of the Hungarian Academy of Sciences and from the NKFI via K-project 138031.

**Data Availability Statement:** The data presented in this study are available upon reasonable request from the authors.

**Acknowledgments:** We deeply thank Philip Adsley for helping with running the RatesMC code.

**Conflicts of Interest:** The authors declare no conflicts of interest.